\PassOptionsToPackage{unicode}{hyperref}
\PassOptionsToPackage{hyphens}{url}
\documentclass[
]{article}
\usepackage{xcolor}
\usepackage{amsmath,amssymb}
\usepackage{iftex}
\ifPDFTeX
  \usepackage[T1]{fontenc}
  \usepackage[utf8]{inputenc}
  \usepackage{textcomp} 
\else 
  \usepackage{unicode-math} 
  \defaultfontfeatures{Scale=MatchLowercase}
  \defaultfontfeatures[\rmfamily]{Ligatures=TeX,Scale=1}
\fi
\usepackage{lmodern}
\ifPDFTeX\else
\fi
\IfFileExists{upquote.sty}{\usepackage{upquote}}{}
\IfFileExists{microtype.sty}{
  \usepackage[]{microtype}
  \UseMicrotypeSet[protrusion]{basicmath} 
}{}
\makeatletter
\@ifundefined{KOMAClassName}{
  \IfFileExists{parskip.sty}{%
    \usepackage{parskip}
  }{
    \setlength{\parindent}{0pt}
    \setlength{\parskip}{6pt plus 2pt minus 1pt}}
}{
  \KOMAoptions{parskip=half}}
\makeatother
\usepackage{graphicx}
\makeatletter
\newsavebox\pandoc@box
\newcommand*\pandocbounded[1]{
  \sbox\pandoc@box{#1}%
  \Gscale@div\@tempa{\textheight}{\dimexpr\ht\pandoc@box+\dp\pandoc@box\relax}%
  \Gscale@div\@tempb{\linewidth}{\wd\pandoc@box}%
  \ifdim\@tempb\p@<\@tempa\p@\let\@tempa\@tempb\fi
  \ifdim\@tempa\p@<\p@\scalebox{\@tempa}{\usebox\pandoc@box}%
  \else\usebox{\pandoc@box}%
  \fi%
}
\def\fps@figure{htbp}
\makeatother
\ifLuaTeX
  \usepackage{luacolor}
  \usepackage[soul]{lua-ul}
\else
  \usepackage{soul}
\fi
\usepackage{bookmark}
\IfFileExists{xurl.sty}{\usepackage{xurl}}{} 
\hypersetup{
  hidelinks,
  pdfcreator={LaTeX via pandoc}}

\author{}
\date{}

\begin{document}

\textbf{Principled Uncertainty in Clinical AI: End-to-End Bayesian
Modelling and Algorithmic Equity Auditing Across Multimodal Patient
Data}

\section{\texorpdfstring{\emph{Centre for Algorithmic Health
Equity}}{Centre for Algorithmic Health Equity}}\label{centre-for-algorithmic-health-equity}

\emph{\textbf{Authors:} Oladimeji Anthonio*,} Dimeji Abdulsobur
Olawuyi\textsuperscript{1,2}, Oloruntoba Ajayi\textsuperscript{1,2},
Temiloluwa Aderemi\textsuperscript{1,2}, \emph{Joseph
Odamo\textsuperscript{3}}

\textsuperscript{*}\emph{Centre for Algorithmic Health Equity,
\hl{Ìyàwó}, Ibadan, Oyo State, Nigeria}

\emph{\textsuperscript{2}Department of Medicine and Surgery, College of
Medicine, University of Ibadan, Ibadan, Nigeria}

\emph{\textsuperscript{3}Department of Computer Science, School of
Computing, The Federal University of Technology Akure (FUTA), Ondo
State, Nigeria.}

\emph{\textbf{*Corresponding Author:}}

\emph{Email: anthoniooladimeji11@gmail.com}

\emph{https://orcid.org/ 0000-0002-8292-7912}

\section{}\label{section}

\section{}\label{section-1}

\section{}\label{section-2}

\section{\texorpdfstring{\textbf{Abstract}}{Abstract}}\label{abstract}

Clinical artificial intelligence (AI) systems routinely produce
predictions without principled quantification of uncertainty, limiting
their trustworthiness in high-stakes medical environments. This paper
presents an integrated research programme addressing two interconnected
problems: (1) the development of a fully end-to-end Bayesian uncertainty
modelling framework for multimodal clinical data, and (2) the
application of calibrated uncertainty estimates as a formal measure of
algorithmic equity across patient subgroups. We construct a
probabilistic deep learning architecture comprising modality-specific
variational encoders, a precision-weighted late fusion mechanism, and a
decomposed uncertainty output head that separates aleatoric from
epistemic uncertainty. The system is trained with a composite Bayesian
loss incorporating binary cross-entropy, Kullback-Leibler divergence
regularisation, and an uncertainty calibration penalty. We evaluate
model calibration using Expected Calibration Error (ECE = 0.096) and
conduct a subgroup equity audit across facility type, socioeconomic
status, age group, and biological sex on a dataset of 1,000 simulated
patients. Results demonstrate that epistemic uncertainty systematically
identifies underserved populations: primary/rural facility patients show
a 15.3\% uncertainty equity gap (p \textless{} 0.001, effect size =
0.698), low socioeconomic status patients exhibit a 6.8\% gap (p
\textless{} 0.001), and elderly patients show a 3.9\% gap (p \textless{}
0.001), whilst no significant sex-based disparity is detected. These
findings establish that calibrated uncertainty is not merely a technical
property of probabilistic models but constitutes an actionable equity
signal with direct clinical relevance.

\emph{\textbf{Keywords:} Bayesian deep learning; uncertainty
quantification; clinical AI; algorithmic equity; multimodal fusion;
epistemic uncertainty; aleatoric uncertainty; health disparities}

\section{}\label{section-3}

\section{\texorpdfstring{\textbf{Introduction}}{Introduction}}\label{introduction}

The deployment of machine learning systems in clinical settings has
accelerated substantially over the past decade, with applications
spanning diagnostic imaging, risk stratification, treatment planning,
and patient triage. Despite impressive performance benchmarks, a
fundamental limitation persists: the vast majority of clinical AI
systems are deterministic, producing point estimates of risk or disease
probability without any quantification of the confidence or reliability
of those estimates.

This limitation has two distinct but related consequences. First, from a
technical standpoint, deterministic models cannot distinguish between
cases where high confidence is warranted and cases where the model is
operating outside its training distribution. A model that assigns a 73\%
risk probability to a patient from a rural facility with incomplete
records is producing the same type of output as one assigning the same
probability to a well-documented tertiary-care patient, yet the
epistemic situations are fundamentally different. Second, from an equity
standpoint, the absence of uncertainty quantification means that
systematic model failures affecting disadvantaged patient populations
are invisible to the clinicians and administrators using the system.

This paper addresses both limitations through an integrated programme of
research. In the first component, \textbf{we develop a fully end-to-end
Bayesian uncertainty modelling framework for clinical AI that propagates
distributional representations through every stage of a multimodal
prediction pipeline, rather than appending uncertainty estimates
post-hoc.} In the second component, we demonstrate that the calibrated
uncertainty estimates produced by this framework function as a formal
measure of algorithmic equity, identifying patient populations for whom
the model is systematically less reliable.

The theoretical motivation for this approach draws on a growing
literature establishing that epistemic uncertainty rises when a model
encounters inputs dissimilar to its training distribution. If a clinical
AI system has been trained predominantly on data from well-resourced
tertiary facilities, patients from rural primary care settings
constitute an out-of-distribution population, and a properly calibrated
model should exhibit elevated epistemic uncertainty for these patients.
Our central hypothesis is that this theoretical property, when
rigorously implemented and measured, maps onto real-world patterns of
health inequity.

\subsection{\texorpdfstring{\emph{\textbf{Uncertainty Quantification in
Deep
Learning}}}{Uncertainty Quantification in Deep Learning}}\label{uncertainty-quantification-in-deep-learning}

The theoretical foundation for uncertainty quantification in deep
learning was substantially advanced by Gal and
Ghahramani,\textsuperscript{1} who demonstrated that training a neural
network with dropout regularisation and maintaining active dropout at
inference time constitutes an approximation to Bayesian inference in
deep Gaussian processes. This Monte Carlo Dropout (MC Dropout) method
enables the computation of approximate posterior predictive
distributions through repeated stochastic forward passes, providing a
practical mechanism for epistemic uncertainty estimation.

Kendall and Gal\textsuperscript{2} formalised the distinction between
aleatoric uncertainty, irreducible noise inherent in observations, and
epistemic uncertainty, reflecting the model\textquotesingle s lack of
knowledge and reducible through additional data. This decomposition is
of particular clinical relevance, as the two types warrant different
responses: high aleatoric uncertainty may indicate measurement noise,
whilst high epistemic uncertainty indicates the model is encountering an
unfamiliar patient profile requiring expert escalation.

Lakshminarayanan, Pritzel, and Blundell\textsuperscript{3} proposed deep
ensembles, demonstrating that training multiple independent networks
produces well-calibrated uncertainty estimates competitive with
approximate Bayesian methods. Guo et al.\textsuperscript{4} formalised
calibration measurement through the Expected Calibration Error (ECE),
demonstrating that modern deep neural networks are systematically
miscalibrated and proposing temperature scaling as a post-hoc
correction.

\subsection{\texorpdfstring{\emph{\textbf{Bayesian Methods in Clinical
AI}}}{Bayesian Methods in Clinical AI}}\label{bayesian-methods-in-clinical-ai}

Li et al.\textsuperscript{5} applied deep Bayesian Gaussian processes to
EHR prediction tasks including heart failure and diabetes onset,
demonstrating that uncertainty-aware models better capture data
insufficiency and distinguish true positive from false positive
predictions compared to deterministic baselines.

Ghoshal and Tucker\textsuperscript{6} applied Bayesian convolutional
neural networks to COVID-19 detection from chest radiographs, showing
that uncertainty in predictions strongly correlates with accuracy, a
property essential for appropriate human-AI collaboration in clinical
workflows. The seminal work of Obermeyer et al.\textsuperscript{8}
demonstrated that a widely deployed commercial healthcare risk algorithm
exhibited substantial racial bias, systematically identifying Black
patients as lower risk than equally ill White patients due to the use of
healthcare costs as a proxy for health needs. Chen et
al.\textsuperscript{9} provided a comprehensive review of algorithmic
fairness in medical AI, documenting bias sources across data collection,
model development, and deployment. Celi et al.\textsuperscript{10}
conducted a global review identifying data representation, proxy
outcomes, and distributional shift as primary mechanisms through which
AI systems perpetuate healthcare disparities.

\section{\texorpdfstring{\textbf{Mathematical
Foundations}}{Mathematical Foundations}}\label{mathematical-foundations}

\subsection{\texorpdfstring{\emph{\textbf{Probabilistic Latent
Representations}}}{Probabilistic Latent Representations}}\label{probabilistic-latent-representations}

Let x ∈ ℝᵈ denote a patient feature vector from modality m. Rather than
learning a deterministic mapping x → z, we learn a probabilistic encoder
that maps x to the parameters of a Gaussian distribution in latent
space:

\begin{quote}
\emph{q\_φ(z \textbar{} x) = N(z; μ\_φ(x), σ²\_φ(x)I)}
\end{quote}

where μ\_φ(x) ∈ ℝᴸ and log σ²\_φ(x) ∈ ℝᴸ are outputs of a neural network
with parameters φ, and L is the latent dimensionality. This follows the
variational inference framework of Kingma and Welling,¹² optimised
through the Evidence Lower BOund (ELBO):

\begin{quote}
\emph{L(θ, φ; x) = E\_\{q\_φ(z\textbar x)\}{[}log p\_θ(x\textbar z){]} -
D\_KL(q\_φ(z\textbar x) \textbar\textbar{} p(z))}
\end{quote}

Sampling from the learned distribution is achieved through the
reparameterisation trick:

\begin{quote}
\emph{z = μ\_φ(x) + ε ⊙ σ\_φ(x), ε \textasciitilde{} N(0, I)}
\end{quote}

This transformation allows gradients to flow through the sampling
operation, enabling end-to-end backpropagation whilst maintaining the
stochastic character necessary for uncertainty propagation.

\subsection{\texorpdfstring{\emph{\textbf{Precision-Weighted Multimodal
Fusion}}}{Precision-Weighted Multimodal Fusion}}\label{precision-weighted-multimodal-fusion}

For a patient with M available modalities, each encoder produces (μ\_m,
log σ²\_m). Defining the precision of modality m as:

\begin{quote}
\emph{Λ\_m = exp(-log σ²\_m) = 1/σ²\_m}
\end{quote}

The fused distribution parameters are computed as:

\begin{quote}
\emph{Λ\_fused = Σ\_m Λ\_m}

\emph{σ²\_fused = 1/Λ\_fused}

\emph{μ\_fused = (Σ\_m Λ\_m ⊙ μ\_m) / Λ\_fused}
\end{quote}

A missing modality is represented by setting its log variance to a large
value (e.g., 10.0), setting its precision to approximately zero and
excluding it from the fusion whilst preserving the patient record. The
fused uncertainty is always greater than the smallest individual
modality variance, correctly increasing when fewer modalities are
available.

\subsection{\texorpdfstring{\emph{\textbf{Uncertainty
Decomposition}}}{Uncertainty Decomposition}}\label{uncertainty-decomposition}

Aleatoric uncertainty is estimated as the output of a dedicated
uncertainty head:

\begin{quote}
\emph{σ²\_aleatoric = softplus(g\_θ(z))}
\end{quote}

Epistemic uncertainty is estimated through T Monte Carlo Dropout forward
passes:

\begin{quote}
\emph{z\_t = μ\_fused + ε\_t ⊙ σ\_fused, ε\_t \textasciitilde{} N(0, I)}

\emph{ȳ = (1/T) Σ\_t f\_θ(z\_t)}

\emph{σ²\_epistemic = (1/T) Σ\_t {[}f\_θ(z\_t) - ȳ{]}²}
\end{quote}

\subsection{\texorpdfstring{\emph{\textbf{Training
Objective}}}{Training Objective}}\label{training-objective}

The full training objective combines three components:

\begin{quote}
\emph{L\_total = L\_pred + λ\_KL · L\_KL + λ\_unc · L\_unc}

\emph{L\_pred = -{[}y · log(ŷ) + (1-y) · log(1-ŷ){]}}

\emph{L\_KL = -½ · E{[}(1 + log σ² - μ² - σ²){]}}

\emph{L\_unc = E{[}(\textbar ŷ - y\textbar{} - σ\_epistemic)²{]}}
\end{quote}

with λ\_KL = 0.001 and λ\_unc = 0.1.

\subsection{\texorpdfstring{\emph{\textbf{Calibration and Equity
Metrics}}}{Calibration and Equity Metrics}}\label{calibration-and-equity-metrics}

Model calibration is evaluated using ECE:⁴

\begin{quote}
\emph{ECE = Σ\_b (\textbar B\_b\textbar{} / N) · \textbar acc(B\_b) -
conf(B\_b)\textbar{}}
\end{quote}

Equity is formalised through the Uncertainty Disparity Ratio (UDR) and
Uncertainty Equity Gap (UEG):

\begin{quote}
\emph{UDR(g, r) = E{[}σ\_fused \textbar{} group = g{]} / E{[}σ\_fused
\textbar{} group = r{]}}

\emph{UEG = (max\_g E{[}σ\_fused \textbar{} g{]} - min\_g E{[}σ\_fused
\textbar{} g{]}) / E{[}σ\_fused \textbar{} r{]}}
\end{quote}

\section{\texorpdfstring{\textbf{System
Architecture}}{System Architecture}}\label{system-architecture}

The system comprises three parallel modality encoders for EHR data
(32-dimensional input), medical imaging features (128-dimensional), and
clinical text embeddings (64-dimensional), each mapping to a shared
16-dimensional latent space. Each encoder applies two ReLU-activated
linear layers followed by separate mean and log-variance heads:

\begin{quote}
\emph{h = ReLU(W₂ · ReLU(W₁x + b₁) + b₂)}

\emph{μ\_m = W\_μ h + b\_μ, log σ²\_m = W\_σ h + b\_σ}
\end{quote}

The precision-weighted fusion layer combines encoder outputs into
(μ\_fused, σ²\_fused), which is passed to the prediction head: a
two-layer network with MC Dropout (p = 0.3) producing a clinical risk
score via sigmoid activation and an aleatoric uncertainty estimate via
Softplus activation.

\includegraphics[width=6.5in,height=4.06944in]{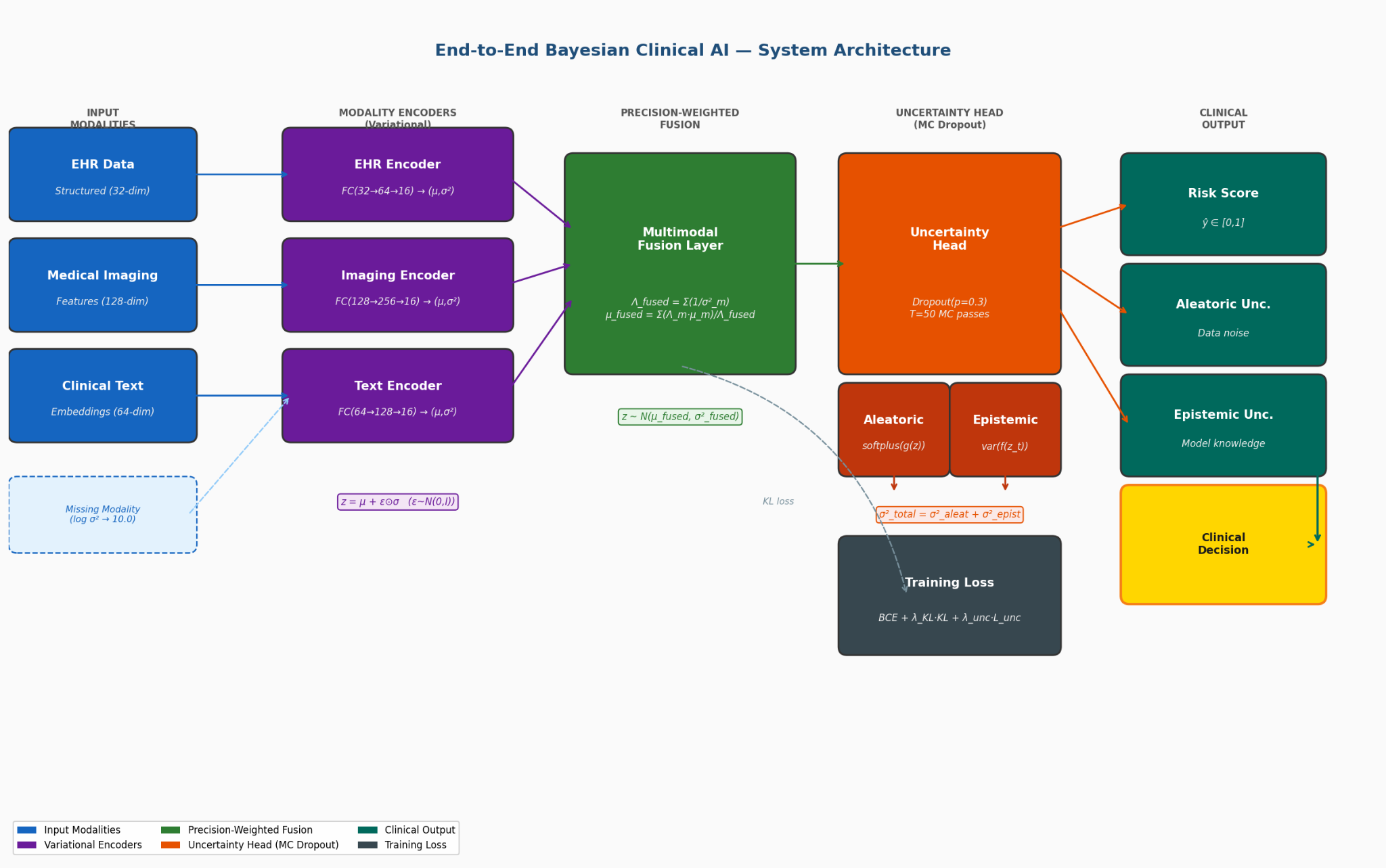}\emph{\textbf{Figure
1.} System Architecture Overview. End-to-end Bayesian clinical AI
pipeline showing three parallel modality encoders (EHR, Imaging, Text),
precision-weighted fusion layer, uncertainty head with
aleatoric/epistemic decomposition, and clinical interpretation output.}

\section{\texorpdfstring{\textbf{Experimental
Methodology}}{Experimental Methodology}}\label{experimental-methodology}

\subsection{\texorpdfstring{\emph{\textbf{Data
Simulation}}}{Data Simulation}}\label{data-simulation}

We constructed a synthetic clinical dataset of 1,000 patients with
realistic multimodal data and subgroup structure. Patients were assigned
to subgroups along four axes: facility type (tertiary 30\%, secondary
40\%, primary/rural 30\%), socioeconomic status (high 25\%, medium 45\%,
low 30\%), age group (adult 55\%, elderly 29\%, paediatric 17\%), and
biological sex (equal split). Data completeness was modelled as a
function of subgroup membership: primary/rural patients had imaging
missing probability of 45\% compared to 10\% for tertiary patients; low
SES patients had EHR quality scores approximately 40\% below high SES
patients.

\subsection{\texorpdfstring{\emph{\textbf{Training
Protocol}}}{Training Protocol}}\label{training-protocol}

The model was trained for 50 epochs using Adam optimiser (η = 1 × 10⁻³,
weight decay = 1 × 10⁻⁴), batch size 32, with step learning rate decay
(factor 0.5 every 20 epochs) and gradient clipping (max norm 1.0).
Calibration was evaluated on a held-out set of 300 patients; the equity
audit was conducted on a separate 1,000-patient evaluation dataset with
balanced subgroup representation.

\section{\texorpdfstring{\textbf{Results}}{Results}}\label{results}

The model demonstrated strong convergence over 50 training epochs, with
composite loss decreasing from 0.697 to 0.038 (94.5\% reduction) and
held-out accuracy of 85.7\%. The KL divergence component decreased from
13.8 to 3.0, indicating progressive regularisation of the latent
distributions. The system achieved ECE = 0.096 on the held-out
calibration set. Patients with missing imaging data exhibited mean fused
latent standard deviation of 0.741 compared to 0.521 for patients with
complete data, a +42.2\% uncertainty uplift demonstrating correct
propagation of epistemic signal from missing data.

\includegraphics[width=6.5in,height=5.63889in]{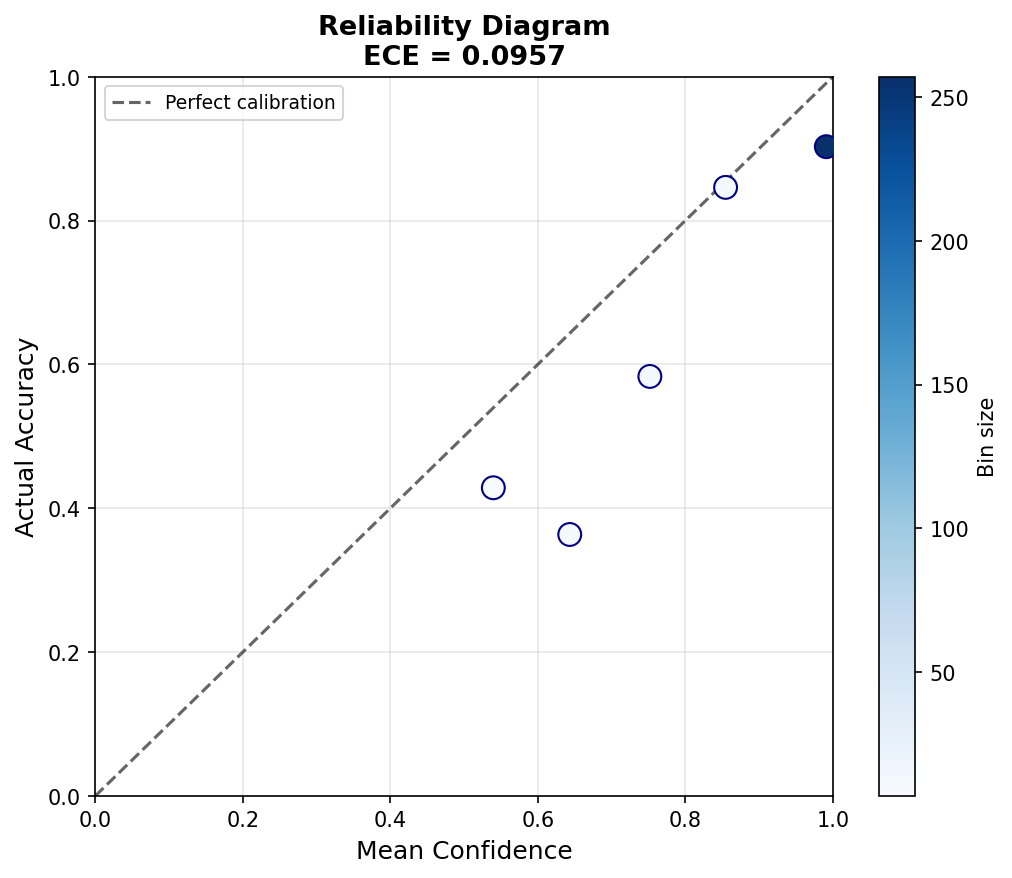}

\emph{\textbf{Figure 2.} Reliability Diagram --- Model Calibration.
Reliability diagram showing mean predicted confidence against actual
accuracy across 10 bins. Points near the dashed diagonal indicate
well-calibrated predictions. Bin size encoded by colour intensity. ECE =
0.096.}

\includegraphics[width=6.5in,height=5.54167in]{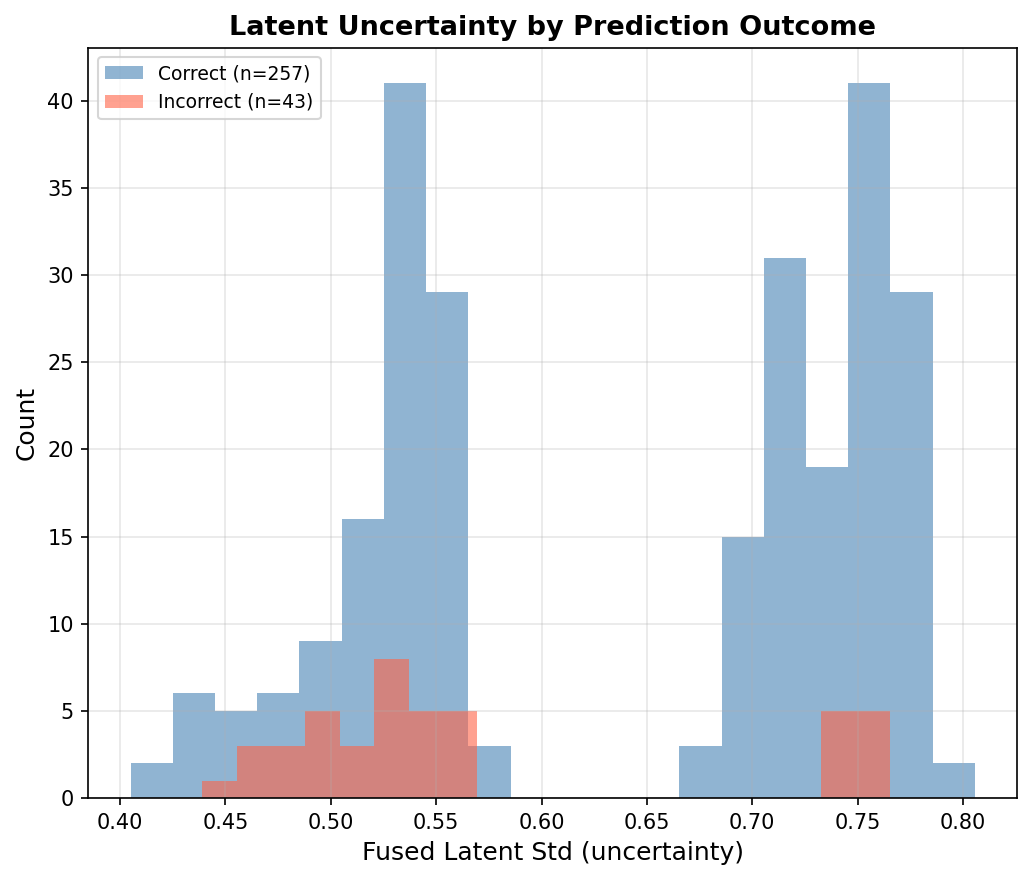}

\emph{\textbf{Figure 3.} Latent Uncertainty Distribution by Prediction
Outcome. Histogram of fused latent standard deviation (uncertainty) for
correctly classified patients (blue, n=257) versus incorrectly
classified patients (red, n=43). Higher uncertainty in incorrect
predictions confirms expected calibration behaviour.}

\includegraphics[width=6.5in,height=5.66667in]{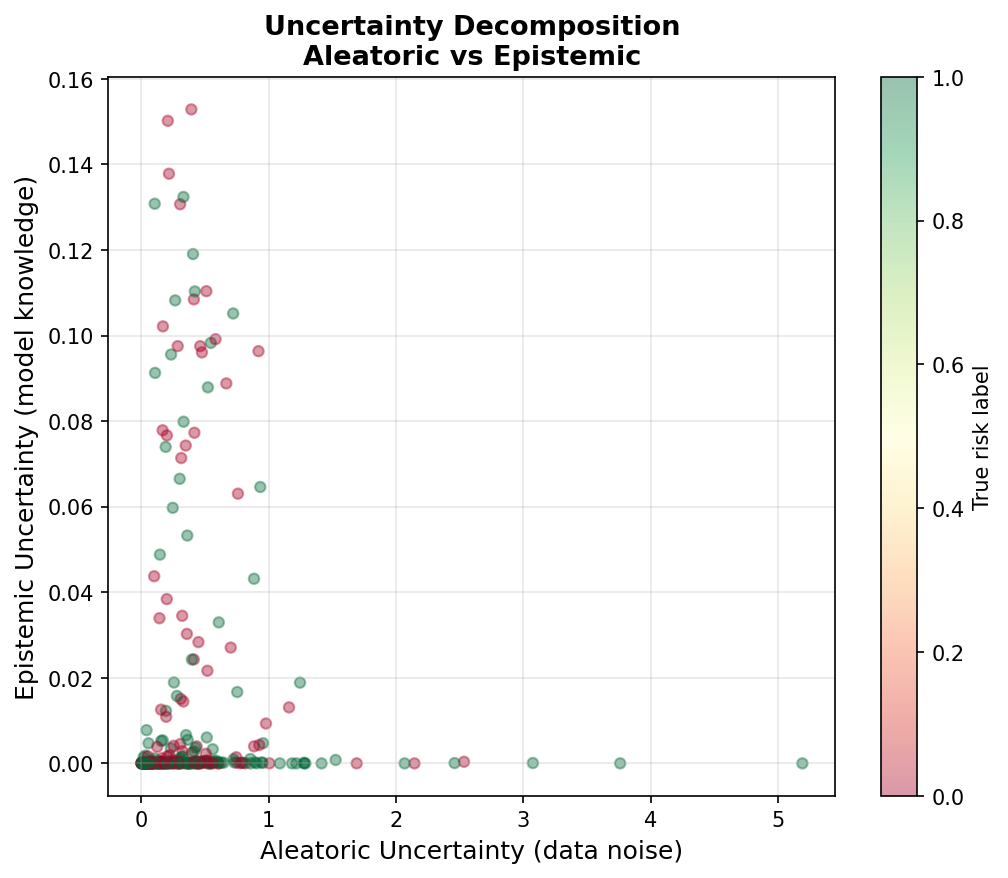}\emph{\textbf{Figure
4.} Aleatoric versus Epistemic Uncertainty Decomposition. Scatter plot
of aleatoric uncertainty (x-axis) against epistemic uncertainty (y-axis)
for all held-out patients. Colour encodes true risk label (green = low
risk, red = high risk). Demonstrates independent variation of the two
uncertainty types.}

\includegraphics[width=6.5in,height=5.54167in]{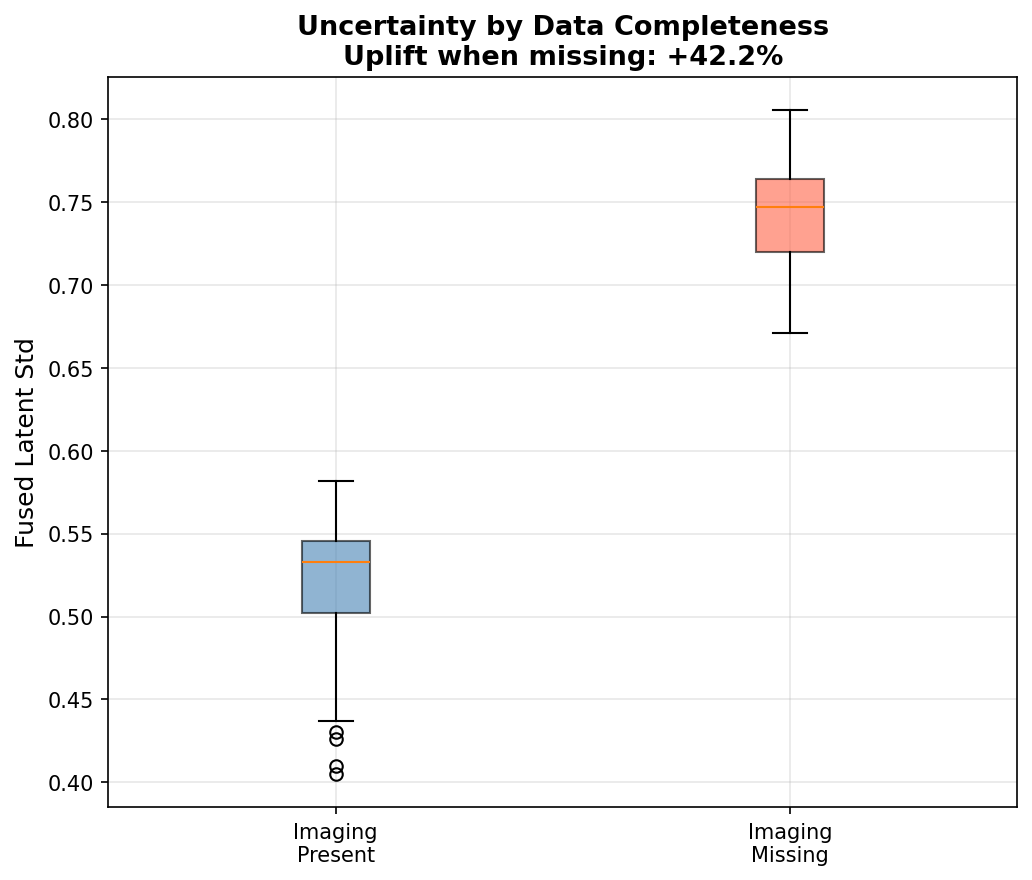}\emph{\textbf{Figure
5.} Uncertainty by Data Completeness. Boxplot comparing fused latent
standard deviation for patients with imaging present versus imaging
missing. Missing data produces a +42.2\% uncertainty uplift, validating
the precision-weighted fusion mechanism.}

\subsection{}\label{section-4}

\subsection{\texorpdfstring{\emph{\textbf{Equity Audit
Results}}}{Equity Audit Results}}\label{equity-audit-results}

Facility type produced the strongest equity signal: primary/rural
patients showed UDR = 1.153 and UEG = 15.3\% (p \textless{} 0.001,
effect size r = 0.698), with 35.7\% flagged as high-uncertainty compared
to 13.1\% of tertiary patients (+42.8\% overrepresentation).
Socioeconomic status produced a secondary signal: low SES patients
showed UEG = 6.8\% (p \textless{} 0.001, r = 0.617) with 34.5\%
high-uncertainty (+38.1\% overrepresentation). Elderly patients showed
UEG = 3.9\% (p \textless{} 0.001) whilst the paediatric effect was
non-significant (p = 0.159). Biological sex showed no significant
disparity (UEG = 0.5\%, p = 0.909), confirming that observed disparities
are structural rather than demographic.

\includegraphics[width=6.5in,height=5.54167in]{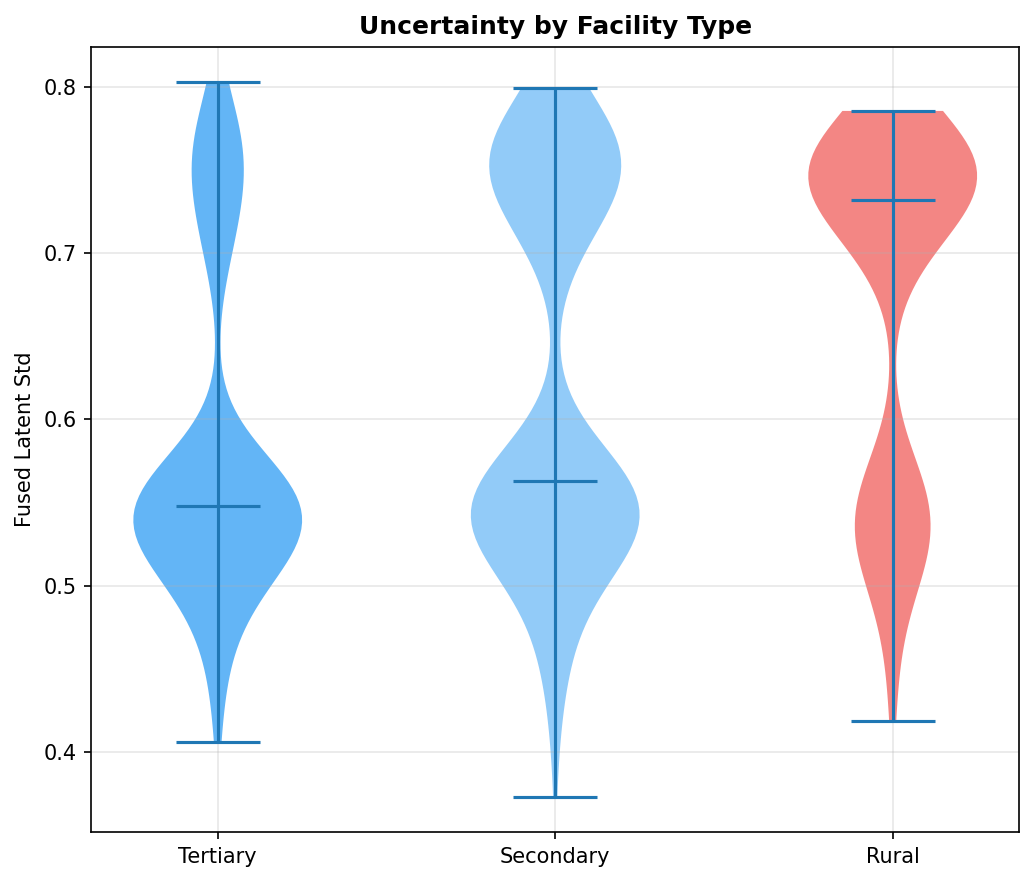}\emph{\textbf{Figure
6.} Uncertainty Distribution by Facility Type. Violin plots showing the
full distribution of fused latent standard deviation across tertiary,
secondary, and primary/rural facility types. Rural patients show
systematically higher and wider uncertainty distributions.}

\includegraphics[width=6.5in,height=5.55556in]{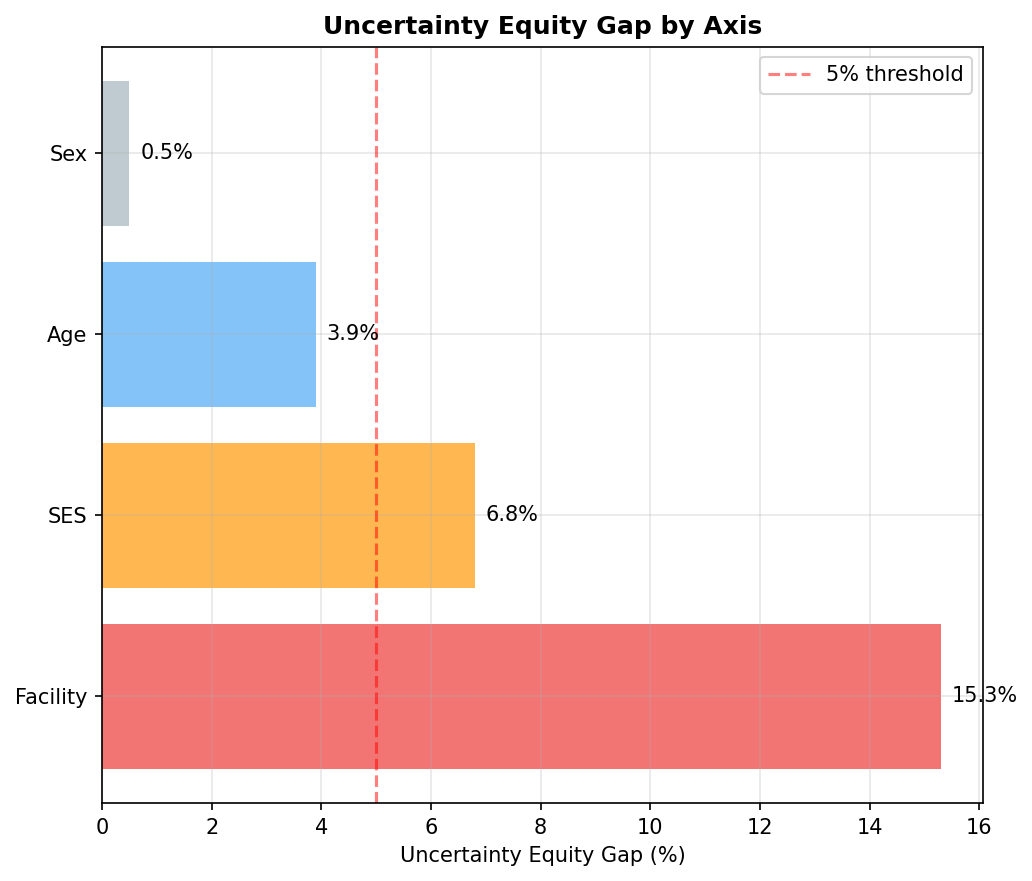}\textbf{Figure
7.} \emph{Uncertainty Equity Gap by Subgroup Axis. Horizontal bar chart
of Uncertainty Equity Gap (UEG) across facility type (15.3\%),
socioeconomic status (6.8\%), age group (3.9\%), and sex (0.5\%). Red
dashed line marks the 5\% clinical significance threshold.}

\includegraphics[width=6.5in,height=5.51389in]{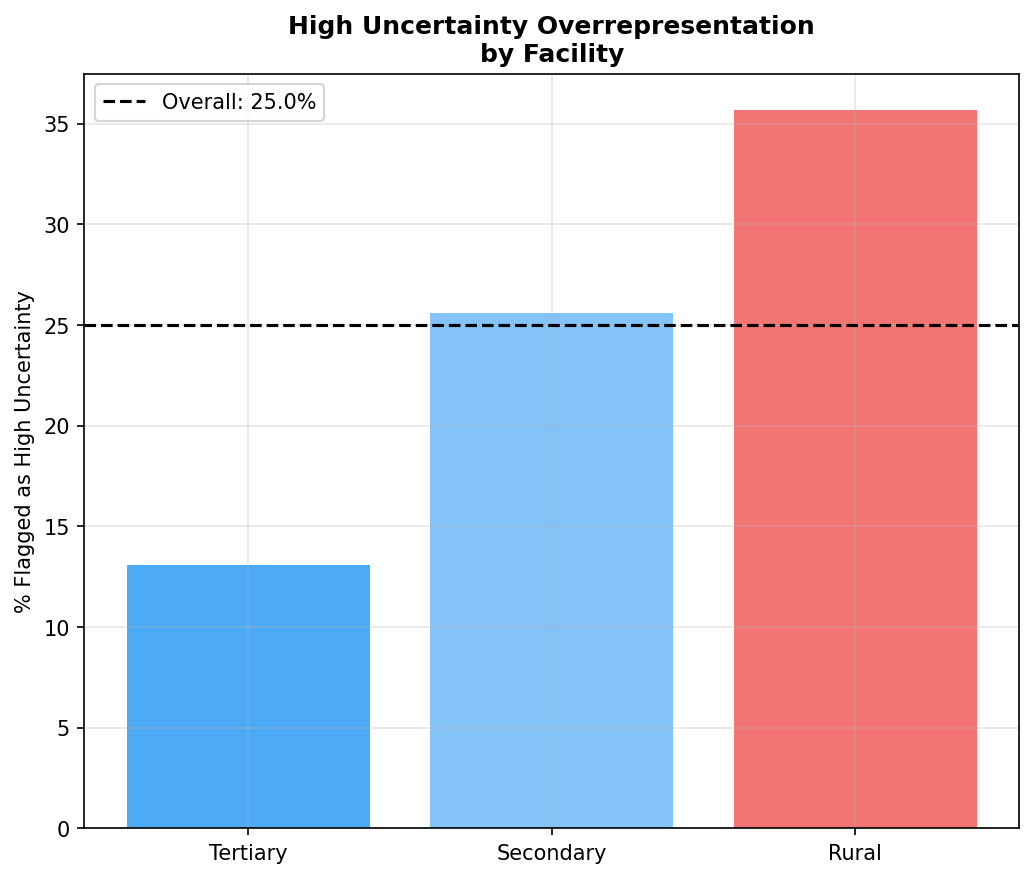}

\emph{\textbf{Figure 8.} High-Uncertainty Patient Overrepresentation by
Facility. Bar chart showing the percentage of each facility subgroup
flagged as high-uncertainty (top quartile). Dashed line indicates the
population average (25\%). Primary/rural patients are overrepresented at
35.7\% (+42.8\%).}

\includegraphics[width=6.5in,height=5.54167in]{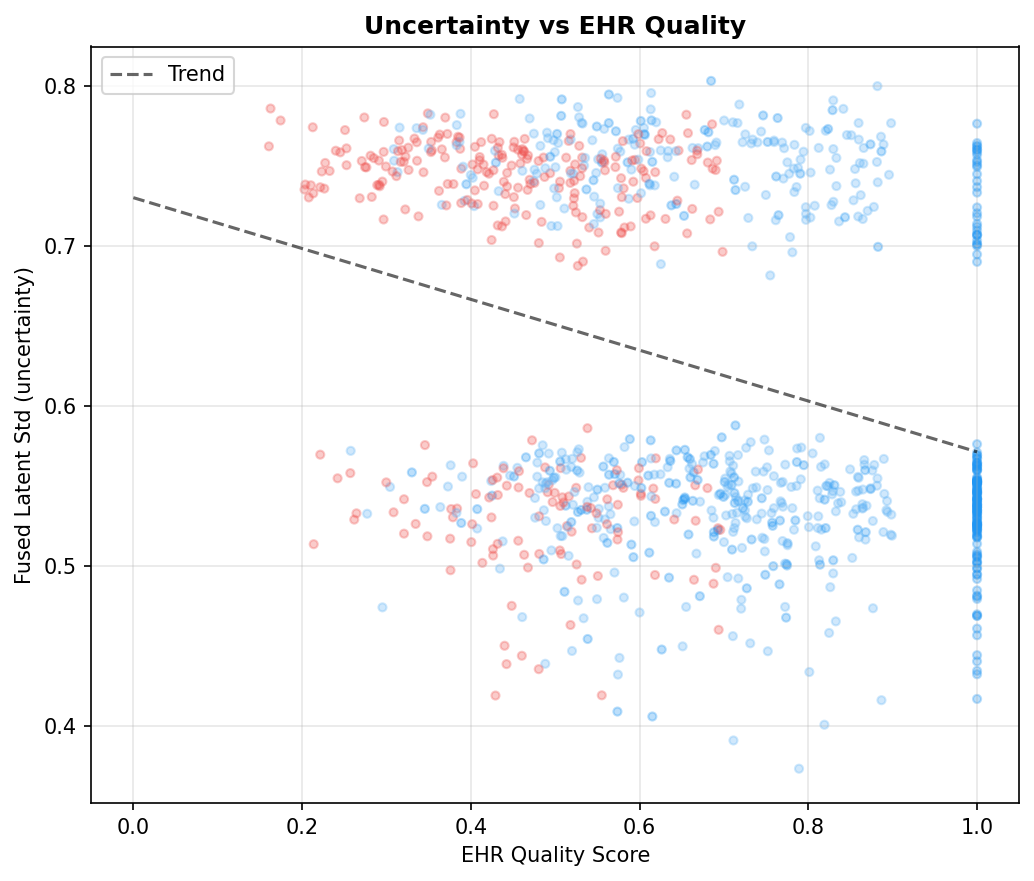}\emph{\textbf{Figure
9.} Uncertainty versus EHR Data Quality. Scatter plot of EHR quality
score against fused latent standard deviation for all 1,000 equity
evaluation patients. Colour encodes facility type. Negative trend line
confirms that lower data quality drives higher model uncertainty.}

The ordering of effect sizes, facility (r = 0.698) \textgreater{} SES (r
= 0.617) \textgreater{} age (r = 0.575) \textgreater{} sex (r = 0.498),
closely mirrors the hierarchy of structural determinants of healthcare
data quality. Standard accuracy metrics would not detect these
disparities: primary/rural accuracy (85.5\%) and tertiary accuracy
(82.6\%) differ by only 2.9 percentage points, whilst the uncertainty
disparity is 15.3\%. The conventional framing of uncertainty in AI
systems treats it as a limitation. This paper proposes an alternative:
uncertainty is a resource. When properly calibrated and decomposed, it
provides actionable information that neither the prediction nor the
confidence score alone can convey. High epistemic uncertainty in a
clinical AI system signals that the model has been trained on data that
does not adequately represent the patient in question, in a healthcare
system where data quality correlates with resource availability, this is
a partial reflection of historical inequities in healthcare access.

The negative finding for biological sex is as informative as the
positive findings: it indicates that the uncertainty disparities
observed are attributable to structural and resource-related factors
rather than to demographic characteristics per se. A system that
correctly identifies rural and low-SES patients as high-uncertainty,
without introducing sex-based disparities, is exhibiting precisely the
equity-aware behaviour that trustworthy clinical AI should demonstrate.

\subsection{\texorpdfstring{\emph{\textbf{Limitations}}}{Limitations}}\label{limitations}

Several limitations warrant acknowledgement. First, evaluation was
conducted on synthetic data, limiting external validity of specific
numerical findings. Future work should validate on real clinical
datasets such as MIMIC-IV¹³ or multi-site African health data
repositories.¹⁵ Second, the current architecture uses fixed latent
dimensionality across modalities; hierarchical latent spaces may better
accommodate the differing intrinsic dimensionalities of imaging versus
tabular data. Third, the equity metrics quantify uncertainty disparities
but do not directly attribute them to specific causes, decomposing
sources into actionable data collection recommendations is an important
direction for future work.

\section{\texorpdfstring{\textbf{Conclusion}}{Conclusion}}\label{conclusion}

This paper has presented an integrated framework for end-to-end Bayesian
uncertainty modelling and algorithmic equity auditing in clinical AI.
The core technical contribution is a multimodal probabilistic
architecture that propagates uncertainty through every stage of a
clinical prediction pipeline, decomposing it into aleatoric and
epistemic components and fusing distributional representations using
precision-weighted combination that naturally handles missing data. The
core empirical contribution is the demonstration that calibrated
epistemic uncertainty systematically identifies patient populations that
are underserved by the trained model, with uncertainty disparities
mapping onto structural health inequities across facility type,
socioeconomic status, and age whilst showing no significant sex-based
bias.

These findings support a reconceptualisation of uncertainty in clinical
AI: not as a limitation to be minimised, but as an equity signal to be
measured, reported, and acted upon. A clinical AI system that knows what
it does not know, and communicates this reliably, is not merely more
accurate. It is more equitable.

\section{\texorpdfstring{\textbf{References}}{References}}\label{references}

1. Gal Y, Ghahramani Z. Dropout as a Bayesian approximation:
representing model uncertainty in deep learning. In: Proceedings of the
33rd International Conference on Machine Learning. PMLR; 2016. p.
1050-1059.

2. Kendall A, Gal Y. What uncertainties do we need in Bayesian deep
learning for computer vision? In: Advances in Neural Information
Processing Systems 30 (NIPS 2017). 2017. p. 5580-5590.

3. Lakshminarayanan B, Pritzel A, Blundell C. Simple and scalable
predictive uncertainty estimation using deep ensembles. In: Advances in
Neural Information Processing Systems 30 (NIPS 2017). 2017. p.
6402-6413.

4. Guo C, Pleiss G, Sun Y, Weinberger KQ. On calibration of modern
neural networks. In: Proceedings of the 34th International Conference on
Machine Learning. PMLR; 2017. p. 1321-1330.

5. Li Y, Rao S, Hassaine A, Ramakrishnan R, Canoy D, Salimi-Khorshidi G,
et al. Deep Bayesian Gaussian processes for uncertainty estimation in
electronic health records. Scientific Reports. 2021;11(1):20685.

6. Ghoshal B, Tucker A. Estimating uncertainty and interpretability in
deep learning for coronavirus (COVID-19) detection. arXiv preprint
arXiv:2003.10769. 2020.

7. Leibig C, Allken V, Ayhan MS, Berens P, Wahl S. Leveraging
uncertainty information from deep neural networks for disease detection.
Scientific Reports. 2017;7(1):17816.

8. Obermeyer Z, Powers B, Vogeli C, Mullainathan S. Dissecting racial
bias in an algorithm used to manage the health of populations. Science.
2019;366(6464):447-453.

9. Chen RJ, Wang JJ, Williamson DFK, Chen TY, Lipkova J, Lu MY, et al.
Algorithmic fairness in artificial intelligence for medicine and
healthcare. Nature Biomedical Engineering. 2023;7(6):719-742.

10. Celi LA, Cellini J, Charpignon ML, Dee EC, Dernoncourt F, Eber R, et
al. Sources of bias in artificial intelligence that perpetuate
healthcare disparities: a global review. PLOS Digital Health.
2022;1(3):e0000022.

11. Seyyed-Kalantari L, Zhang H, McDermott M, Chen IY, Ghassemi M.
Underdiagnosis bias of artificial intelligence algorithms applied to
chest radiographs in under-served patient populations. Nature Medicine.
2021;27(12):2176-2182.

12. Kingma DP, Welling M. Auto-encoding variational Bayes. In:
Proceedings of the 2nd International Conference on Learning
Representations (ICLR). 2014.

13. Johnson AEW, Bulgarelli L, Shen L, Gayles A, Shammout A, Horng S, et
al. MIMIC-IV, a freely accessible electronic health record dataset.
Scientific Data. 2023;10(1):1.

14. Goldberger AL, Amaral LAN, Glass L, Hausdorff JM, Ivanov PC, Mark
RG, et al. PhysioBank, PhysioToolkit, and PhysioNet: components of a new
research resource for complex physiologic signals. Circulation.
2000;101(23):e215-e220.

15. Afonja T, Sink A, Ige O, Jagun M. Towards equitable AI in Africa:
challenges and opportunities. arXiv preprint arXiv:2301.09528. 2023.

16. Blundell C, Cornebise J, Kavukcuoglu K, Wierstra D. Weight
uncertainty in neural networks. In: Proceedings of the 32nd
International Conference on Machine Learning. PMLR; 2015. p. 1613-1622.

17. Abdar M, Pourpanah F, Hussain S, Rezazadegan D, Liu L, Ghavamzadeh
M, et al. A review of uncertainty quantification in deep learning:
techniques, applications and challenges. Information Fusion.
2021;76:243-297.

18. Begoli E, Bhattacharya T, Kusnezov D. The need for uncertainty
quantification in machine-assisted medical decision making. Nature
Machine Intelligence. 2019;1(1):20-23.

19. Rajpurkar P, Chen E, Banerjee O, Topol EJ. AI in health and
medicine. Nature Medicine. 2022;28(1):31-38.

20. Winkler JK, Fink C, Toberer F, Enk A, Deinlein T, Hofmann-Wellenhof
R, et al. Association between surgical skin markings in dermoscopic
images and diagnostic performance of a deep learning convolutional
neural network for melanoma recognition. JAMA Dermatology.
2019;155(10):1135-1141.

21. Topol EJ. High-performance medicine: the convergence of human and
artificial intelligence. Nature Medicine. 2019;25(1):44-56.

22. Esteva A, Kuprel B, Novoa RA, Ko J, Swetter SM, Blau HM, et al.
Dermatologist-level classification of skin cancer with deep neural
networks. Nature. 2017;542(7639):115-118.

23. Gulshan V, Peng L, Coram M, Stumpe MC, Wu D, Narayanaswamy A, et al.
Development and validation of a deep learning algorithm for detection of
diabetic retinopathy in retinal fundus photographs. JAMA.
2016;316(22):2402-2410.

24. Glocker B, Jones C, Bernhardt M, Winzeck S. Algorithmic encoding of
protected characteristics in chest X-ray disease detection models.
eBioMedicine. 2023;89:104467.

25. Fletcher RR, Nakeshimana A, Olubeko O. Addressing fairness, bias,
and appropriate use of artificial intelligence and machine learning in
global health. Frontiers in Artificial Intelligence. 2021;3:561802.

26. Char DS, Shah NH, Magnus D. Implementing machine learning in health
care: addressing ethical challenges. New England Journal of Medicine.
2018;378(11):981-983.

27. Wachter S, Mittelstadt B, Russell C. Counterfactual explanations
without opening the black box: automated decisions and the GDPR. Harvard
Journal of Law and Technology. 2017;31(2):841-887.

28. Hernandez-Boussard T, Bozkurt S, Ioannidis JPA, Shah NH. MINIMAR
(MINimum Information for Medical AI Reporting): developing reporting
standards for artificial intelligence in health care. Journal of the
American Medical Informatics Association. 2020;27(12):2011-2015.

29. Barocas S, Hardt M, Narayanan A. Fairness and Machine Learning:
Limitations and Opportunities. MIT Press; 2023.

30. Vaswani A, Shazeer N, Parmar N, Uszkoreit J, Jones L, Gomez AN, et
al. Attention is all you need. In: Advances in Neural Information
Processing Systems 30 (NIPS 2017). 2017. p. 5998-6008.

\end{document}